\documentclass[aps,prd,twocolumn,superscriptaddress,nofootinbib]{revtex4-1}
\usepackage{amsfonts}
\usepackage{psfrag}
\usepackage{graphicx}
\usepackage{cancel}
\usepackage{amsmath}
\usepackage{natbib}
\usepackage{xr}
\usepackage{hyperref}
\usepackage{cleveref}
\usepackage{verbatim}
\usepackage{multirow}

\usepackage{booktabs,dcolumn}

\newcolumntype{d}[1]{D{.}{.}{#1}}

\RequirePackage[normalem]{ulem} %DIF PREAMBLE
\RequirePackage{color}\definecolor{RED}{rgb}{1,0,0}\definecolor{BLUE}{rgb}{0,0,1} %DIF PREAMBLE
\providecommand{\DIFaddbegin}{} %DIF PREAMBLE
\providecommand{\DIFaddend}{} %DIF PREAMBLE
\providecommand{\DIFdelbegin}{} %DIF PREAMBLE
\providecommand{\DIFdelend}{} %DIF PREAMBLE
\providecommand{\DIFaddbeginFL}{} %DIF PREAMBLE
\providecommand{\DIFaddendFL}{} %DIF PREAMBLE
\providecommand{\DIFdelbeginFL}{} %DIF PREAMBLE
\providecommand{\DIFdelendFL}{} %DIF PREAMBLE
\newcommand{\DIFscaledelfig}{0.5}
\RequirePackage{settobox} %DIF PREAMBLE
\RequirePackage{letltxmacro} %DIF PREAMBLE
\newsavebox{\DIFdelgraphicsbox} %DIF PREAMBLE
\newlength{\DIFdelgraphicswidth} %DIF PREAMBLE
\newlength{\DIFdelgraphicsheight} %DIF PREAMBLE
\LetLtxMacro{\DIFOincludegraphics}{\includegraphics} %DIF PREAMBLE
\newcommand{\DIFaddincludegraphics}[2][]{{\color{blue}\fbox{\DIFOincludegraphics[#1]{#2}}}} %DIF PREAMBLE
\newcommand{\DIFdelincludegraphics}[2][]{% %DIF PREAMBLE
\sbox{\DIFdelgraphicsbox}{\DIFOincludegraphics[#1]{#2}}% %DIF PREAMBLE
\settoboxwidth{\DIFdelgraphicswidth}{\DIFdelgraphicsbox} %DIF PREAMBLE
\settoboxtotalheight{\DIFdelgraphicsheight}{\DIFdelgraphicsbox} %DIF PREAMBLE
\scalebox{\DIFscaledelfig}{% %DIF PREAMBLE
\parbox[b]{\DIFdelgraphicswidth}{\usebox{\DIFdelgraphicsbox}\\[-\baselineskip] \rule{\DIFdelgraphicswidth}{0em}}\llap{\resizebox{\DIFdelgraphicswidth}{\DIFdelgraphicsheight}{% %DIF PREAMBLE
\setlength{\unitlength}{\DIFdelgraphicswidth}% %DIF PREAMBLE
\begin{picture}(1,1)% %DIF PREAMBLE
\thicklines\linethickness{2pt} %DIF PREAMBLE
{\color[rgb]{1,0,0}\put(0,0){\framebox(1,1){}}}% %DIF PREAMBLE
{\color[rgb]{1,0,0}\put(0,0){\line( 1,1){1}}}% %DIF PREAMBLE
{\color[rgb]{1,0,0}\put(0,1){\line(1,-1){1}}}% %DIF PREAMBLE
\end{picture}% %DIF PREAMBLE
}\hspace*{3pt}}} %DIF PREAMBLE
} %DIF PREAMBLE
\LetLtxMacro{\DIFOaddbegin}{\DIFaddbegin} %DIF PREAMBLE
\LetLtxMacro{\DIFOaddend}{\DIFaddend} %DIF PREAMBLE
\LetLtxMacro{\DIFOdelbegin}{\DIFdelbegin} %DIF PREAMBLE
\LetLtxMacro{\DIFOdelend}{\DIFdelend} %DIF PREAMBLE
\DeclareRobustCommand{\DIFaddbegin}{\DIFOaddbegin \let\includegraphics\DIFaddincludegraphics} %DIF PREAMBLE
\DeclareRobustCommand{\DIFaddend}{\DIFOaddend \let\includegraphics\DIFOincludegraphics} %DIF PREAMBLE
\DeclareRobustCommand{\DIFdelbegin}{\DIFOdelbegin \let\includegraphics\DIFdelincludegraphics} %DIF PREAMBLE
\DeclareRobustCommand{\DIFdelend}{\DIFOaddend \let\includegraphics\DIFOincludegraphics} %DIF PREAMBLE
\LetLtxMacro{\DIFOaddbeginFL}{\DIFaddbeginFL} %DIF PREAMBLE
\LetLtxMacro{\DIFOaddendFL}{\DIFaddendFL} %DIF PREAMBLE
\LetLtxMacro{\DIFOdelbeginFL}{\DIFdelbeginFL} %DIF PREAMBLE
\LetLtxMacro{\DIFOdelendFL}{\DIFdelendFL} %DIF PREAMBLE
\DeclareRobustCommand{\DIFaddbeginFL}{\DIFOaddbeginFL \let\includegraphics\DIFaddincludegraphics} %DIF PREAMBLE
\DeclareRobustCommand{\DIFaddendFL}{\DIFOaddendFL \let\includegraphics\DIFOincludegraphics} %DIF PREAMBLE
\DeclareRobustCommand{\DIFdelbeginFL}{\DIFOdelbeginFL \let\includegraphics\DIFdelincludegraphics} %DIF PREAMBLE
\DeclareRobustCommand{\DIFdelendFL}{\DIFOaddendFL \let\includegraphics\DIFOincludegraphics} %DIF PREAMBLE
\newcommand{\e}{\mathrm{e}}
\newcommand{\D}{\mathrm{d}}

\begin{document}

\title{Analysis of Bose-Einstein condensation times for self-interacting scalar dark matter}

\author{Kay Kirkpatrick}
\email{kkirkpat@illinois.edu}
\affiliation{Department of Mathematics,
University of Illinois at Urbana-Champaign
Urbana, IL 61801}
\affiliation{Department of Physics,
University of Illinois at Urbana-Champaign
Urbana, IL 61801}
\author{Anthony E. Mirasola}
\email{aem8@illinois.edu}
\affiliation{Department of Physics,
University of Illinois at Urbana-Champaign
Urbana, IL 61801}
\author{Chanda Prescod-Weinstein}
\email{Chanda.Prescod-Weinstein@unh.edu}
\affiliation{Department of Physics \& Astronomy, University of New Hampshire, Durham, NH 03824}

\begin{abstract}

%In this paper, we extend prior work investigating the condensation time of self-interacting axion-like particles in a gravitational well. Prior work \cite{Relax} showed that the Wigner formalism is the correct analytic approach to describing a condensing scalar field. In this work, we use this formalism to affirm that quartic self-interactions will take longer than necessary to support the time scales associated with structure formation, making gravity a necessary part of the process to bring axion dark matter into a solitonic form. Here we build on prior work by re-calculating the time scale associated with self-interactions, making a prior calculation more exact. We show analytically that the time for condensation will scale with the square of the self-interaction coupling strength. This is consistent with recent numerical estimates, and it affirms that the Wigner formalism is a helpful analytic framework to describe this scenario, which provides a check on computational work that has the potential to introduce numerical artifacts. 

We investigate the condensation time of self-interacting axion-like particles in a gravitational well, extending the prior work~\cite{Relax} which showed that the Wigner formalism is a good analytic approach to describe a condensing scalar field. In the present work, we use this formalism to affirm that $\phi^4$ self-interactions will take longer than necessary to support the time scales associated with structure formation, making gravity a necessary part of the process to bring axion dark matter into a solitonic form. Here we show that when the axions' virial velocity is taken into account, the time scale associated with self-interactions will scale as $\lambda^2$. This is consistent with recent numerical estimates, and it confirms that the Wigner formalism described in prior work~\cite{Relax} is a helpful analytic framework to check computational work for potential numerical artifacts.

%

%\textit{This preprint is available as arXiv:2007.07438 and is highly recommended for publication in Physical Review D.}
\end{abstract}

\maketitle

\section{Introduction}
%The composition of dark matter is one of the most longstanding problems in cosmology. The dominant model, known as Lambda Cold Dark Matter ($\Lambda$CDM), proposes that the dark matter is cold and has a low velocity dispersion. It has been successful at cosmological distance scales \cite{PLANCK2018}. However, at galactic distance scales and smaller ($\lesssim 10$ kpc) it has a number of problems. At these scales, the predicted density profiles disagree with observations and a higher abundance of dwarf galaxies is predicted than is observed \cite{WeinbergControversies2015,CDMproblems,TooBigToFail}. While there are several proposed solutions to these problems \cite{AvilaReese2001,Kamionkowski2000,Spergel2000,Governato2010,BuckleyPeter2018}, an attractive proposal considers the quantum properties of the dark matter particles. In this case, the large-scale predictions remain the same as in $\Lambda$CDM, but on scales less than the de~Broglie wavelength the predictions change.

Axions and axion-like particles have become some of the most well-motivated candidates for the dark matter. Axions across a wide mass range could have the correct abundance to compose the dark matter \cite{Preskill1983,FischlerDine1983,AbbottSikivie,Kim2010}. Additionally axions can address a number of longstanding problems such as the strong CP problem in QCD, and they can play a role in inflation and as dark energy \cite{PecceiQuinn,Weinberg1978,Wilczek1978,DineFischlerSrednicki,Turner1986,SikivieNotes,AxionCosmology}. 

As a dark matter candidate, axions can solve problems with small-scale structure formation which are difficult to explain under the dominant weakly interacting massive particle (WIMP) model \cite{WeinbergControversies2015,CDMproblems,TooBigToFail}. While on cosmological distance scales the predictions of axion models agree with the predictions of WIMPs, axions behave differently on the length scale of their de Broglie wavelength. This not only addresses the small-scale structure problems of $\Lambda$CDM, but it could also lead to novel astrophysical structures which could one day be observed, for example through gravitational wave signatures from collisions with other objects \cite{DietrichCollisions,DietrichCollisionsBlackHole,XiaolongCollisions}.

One such structure is gravitationally bound solitons which are believed to form in regions of axion overdensity \cite{Lee1992,Jetzer1992,Kolb1993,SelfInteractionSign,SemikozTkachev,Khlebnikov2000,Sikivie2009,Erken2012}. These solitons are often called Bose stars since the axion field condenses into a Bose Einstein condensate (BEC). While there have been many studies of these objects, especially their equilibrium properties, there are still questions about their formation process. Since relaxation into the BEC phase is driven by the interactions between particles, and the two interactions (gravitational and self-interactions) at play in the axion field are extremely weak, this process can take an exceedingly long time. Recently, it has become clear that gravity alone is sufficient to drive the condensation process \cite{SemikozTkachev,Khlebnikov2000,Sikivie2009,Schive2014,Levkov2018,AminMocz,XiaolongSimulations}, but there are still questions as to what role, if any, the axion self-interaction plays in this process, particularly in axion species where these self-interactions are stronger than in QCD axions.
     
In this work we evaluate the effect of the initial conditions of the axions in virialized minclusters on the non-equilibrium condensation process. In Ref.\ \cite{Relax}, we previously showed that the evolution equation of the Wigner function describes the relaxation process, and found that gravitational scattering alone was sufficient to cause condensation. In this paper, in the opposite regime where self-interactions drive condensation, we 
%find that the virialized velocities of the axions in miniclusters greatly lengthens the timescale for Bose stars to form through the axion's self-interaction, 
analytically derive results that confirm previous estimates \cite{Sikivie2009}. Specifically, 
%while our previous work showed that the relaxation rate for condensation was proportional to $\lambda$,
using the analytic framework described in~\cite{Relax} we show here that when the limit on the axions' energy set by the virial velocity is taken into account, the relaxation rate scales with $\lambda^2$. These new results update our prior ones.

This paper is organized as follows. In Sec.\ II, we evaluate the role of self interactions in the relaxation by expanding the equation of motion for the axion field's Wigner function. In Sec.\ III, we conclude and compare the impact of self-interactions and gravity on the condensation process.

\section{Wigner function evolution}
\subsection{Initial conditions in miniclusters}
The QCD axion undergoes a phase transition when the universe cools below the Peccei-Quinn symmetry breaking scale, $f_a\sim 10^{12}$ GeV. Depending on the energy scale of inflation, this can either occur during inflation or after it has ended. In the post-inflationary scenario, which we consider here, $f_a$ is lower than the energy scale of inflation, and the result is the formation of gravitationally bound structures through the Kibble mechanism \cite{AxionCosmology}. 
Symmetry breaking causes the axion field to take nonzero values. However, these values can differ in causally disconnected regions of the universe, resulting in a field with random, $\mathcal O(1)$ fluctuations on a distance scale set by the size of the Hubble volume during symmetry breaking. These fluctuations can decouple from the background Hubble expansion to form gravitationally-bound structures we refer to as axion miniclusters \cite{HoganRees,Kolb1993,Nelson2018,Kibble1976}.

The size of the miniclusters depends on the symmetry breaking scale $f_a$ and hence implicitly on the mass of the axion. For the QCD axion, the typical minicluster size is $R\sim 10^5$ km \cite{MiniclusterSpectrum}. Ultralight axions and other scalar dark matter can form dark matter clusters and halos through the same mechanism, though the size of these clusters depends on the model in question \cite{AxionCosmology}. 

The typical velocity of the field is also determined by these initial conditions. Since the miniclusters are gravitationally bound objects, the velocity is determined by the virial theorem, which implies
\begin{equation}
    v= \frac{4\pi G m n R^2}{3}.
\end{equation}
For the QCD axion, the virial velocity within the miniclusters is $v\sim 10^{-10} c$.

The result of this process is that the axion field $\psi(\mathbf x,t)$ is only supported within a region of radius $R$. Similarly, the wavenumbers which are occupied by the axion field are limited. The de Broglie wavelength corresponding to the virial velocity provides an upper bound, and the result is that the field is occupied only for wavenumbers in the range
\begin{equation}
   k \lesssim m v/\hbar .
   \label{eq:Wavenumber}
\end{equation}

\subsection{Evolution due to self-interactions}
The virial velocity $v$ is much smaller than $c$, so the field is non-relativistic. In the non-relativistic regime, the real scalar field $\phi$ that describes the axion can be replaced with a complex scalar field $\psi$ through
\begin{equation}
    \phi=\frac{1}{\sqrt{2m}}\left[\e^{-imt}\psi + \e^{imt}\psi^* \right].
    \label{eq:nonrel}
\end{equation}
(Throughout the paper, we suppress the time-dependence of all fields and write, e.g., $\psi(t)=\psi$.)
Because it is non-relativistic the field $\psi$ evolves under Newtonian potential $U$ due to the field's self-gravity. It also experiences self-interactions arising from the QCD axion potential,
\begin{equation}
     V(\phi)=m^2 f_a ^2\left[1- \cos\frac{\phi}{f_a} \right].
    \label{axionpotential}
\end{equation}
In the non-relativistic limit the field is much lower than the symmetry breaking scale, and the cosine potential can be expanded as
\begin{equation}
    V(\phi)=\frac{1}{2}m^2\phi^2 + \lambda \phi^4,
    \label{eq:fourthorder}
\end{equation}
with
\begin{equation}
    \lambda =-\frac{1}{4!} \frac{m^2}{f_a ^2}.
\end{equation}
Ultralight axions and other scalar field dark matter also experience quartic self-interactions, though the relation between the coupling constant $\lambda$ and the other parameters in the model may differ in those cases. 

The evolution of the non-relativistic field $\psi$ under both these interactions is determined by the Gross-Pitaevskii-Poisson (GPP) equations,
\begin{equation}\begin{aligned}
    i\partial_t \psi &= -\Delta \psi/2m + U\psi +g |\psi|^2 \psi \\
    \Delta U &= 4\pi G m^2 (|\psi|^2 - n_0),
    \label{eq:Poisson}
\end{aligned}\end{equation}
where the background density $n_0$ is removed as a source in the Poisson equation \cite{SelfInteractionSign,Schive2014}. Here we have introduced a dimensionful coupling constant
\begin{equation}
    g=\frac{\lambda}{8m^2}.
\end{equation}
It is useful to consider the evolution of the Fourier-transformed field in momentum space, which we will write as $\psi_\mathbf p$. To obtain the momentum-space equation of motion for  $\psi_{\mathbf p}$, one differentiates the Hamiltonian with respect to $\psi^*_{\mathbf p}$. 
The Hamiltonian is
\begin{equation}
    H = \omega_\mathbf p |\psi_\mathbf p|^2 + H_\mathrm{int}
\end{equation}
where the kinetic energy is
\begin{equation}
    \omega_\mathbf p = \frac{|\mathbf p|^2}{2m},
\end{equation}
and the interaction Hamiltonian is,
\begin{equation}
    H_\mathrm{int} = \int (\prod_{i=1} ^3\D \mathbf q_i) V_{\mathbf p \mathbf q_1 \mathbf q_2 \mathbf q_3} \psi^* _{\mathbf p} \psi^* _{\mathbf q_1} \psi_{\mathbf q_2} \psi_{\mathbf q_3}
\end{equation}
\begin{equation}
    V_{\mathbf p \mathbf q_1 \mathbf q_2 \mathbf q_3} = g \delta(\mathbf p+ \mathbf q_1-\mathbf q_2-\mathbf q_3).
\end{equation}
In general, there is an additional interaction term for the gravitational interactions between particles. In this paper we study the effect of the self-interactions, so we turn gravity off and compare condensation due to gravity with condensation due to self-interactions in the discussion.

The momentum space equation of motion is obtained by differentiation 
\begin{equation}
    i\frac{\partial \psi_\mathbf{p}}{\partial t} = \omega_\mathbf p \psi_\mathbf{p} + S_\mathbf{p},
    \label{eq:GPE}
\end{equation}
where the self-interaction term is
\begin{equation}\begin{aligned}
    S_\mathbf p &= \frac{\delta H_\mathrm{int}}{\delta \psi^* _\mathbf{p}} \\
    &=g \int (\prod_{i=1} ^3\D \mathbf q_i) \delta(\mathbf{p+q_1-q_2-q_3})\psi^* _\mathbf{q_1}\psi _\mathbf{q_2}\psi _\mathbf{q_3}
\end{aligned}\end{equation}

The Wigner function (sometimes called the Wigner quasiprobability distribution), which characterizes a statistical ensemble of fields, is
\begin{equation}\begin{aligned}
    f(\mathbf{x,p}) &= 
    \int \D\mathbf y\, \e^{-i\mathbf p\cdot \mathbf y} \langle\psi^*(\mathbf{x+y}/2)\psi(\mathbf{x-y}/2)\rangle \\
    &=\int \D\mathbf q \,\e^{i\mathbf q\cdot \mathbf x} \langle\psi^* _{\mathbf p +\frac{\mathbf  q}{2}}\psi_{\mathbf p -\frac{\mathbf  q}{2}}\rangle.
    \label{eq:Wigner}
\end{aligned}\end{equation}
The angle brackets $\langle\cdot\rangle$ are an ensemble average over the distribution, which we take to be a distribution of Gaussian fields with random phases. The Wigner function describes the density of the axions in phase space. Unlike a distribution of classical particles, it can take on negative values in phase space regions with area less than $\hbar$. The phenomenon is the result of interference between waves.

The equation of motion for the Wigner function can be obtained by applying Eq. (\ref{eq:GPE}) to each field in the integrand of Eq. (\ref{eq:Wigner}). We obtain,
\begin{equation}\begin{aligned}
       &\frac{\D f(\mathbf{x,p})}{\D t} = \\ &\qquad i\int\D\mathbf q \e^{i\mathbf {q\cdot x}} \left[(\omega_{\mathbf p +\frac{\mathbf  q}{2}} - \omega_{\mathbf p -\frac{\mathbf  q}{2}}) \langle\psi^* _{\mathbf p +\frac{\mathbf  q}{2}}\psi _{\mathbf p -\frac{\mathbf  q}{2}}\rangle\right. \\
       &\qquad\qquad\qquad\quad\left. +\langle S^* _{\mathbf p +\frac{\mathbf  q}{2}}\psi _{\mathbf p -\frac{\mathbf  q}{2}}\rangle -\langle \psi^* _{\mathbf p +\frac{\mathbf  q}{2}}S_{\mathbf p -\frac{\mathbf  q}{2}}\rangle \right].
\end{aligned}\end{equation}
So we have
\begin{equation}\begin{aligned}
    &\frac{\D f(\mathbf{x,p})}{\D t} =\\
    &\quad i\int\D\mathbf q \e^{i\mathbf {q\cdot x}}\frac{\mathbf{p\cdot q}}{m}\langle \psi^* _{\mathbf p+\frac{\mathbf q}{2}} \psi_{\mathbf p -\frac{\mathbf q}{2}}\rangle
    + i\int \D\mathbf q (\prod_{i=1} ^3 \D\mathbf q_i) \e^{i\mathbf{q\cdot x}}\\
    &\quad \times g \left[\delta(\mathbf p +\frac{\mathbf q}{2}+\mathbf q_1 -\mathbf q_2 -\mathbf q_3) \langle \psi _{\mathbf q_1} \psi^* _{\mathbf q_2}\psi^* _{\mathbf q_3} \psi_{\mathbf p-\frac{\mathbf q}{2}}\rangle \right.\\
    &\qquad\,\,\left.- \delta(\mathbf p  -\frac{\mathbf q}{2}+\mathbf{q_1} -\mathbf{q_2} -\mathbf{q_3}) \langle\psi^*_{\mathbf p+\frac{\mathbf q}{2}} \psi^* _{\mathbf q_1} \psi_{\mathbf q_2} \psi_{\mathbf q_3}\rangle \right].
\end{aligned}\end{equation}
Assuming the fields are spatially homogeneous, we can take $\mathbf x = 0$, and the first term vanishes. Introducing notation for the four-point correlator,
\begin{equation}
    J_{\mathbf p_1 \mathbf p_2 \mathbf p_3 \mathbf p_4} \delta(\mathbf p_1+\mathbf p_2-\mathbf p_3-\mathbf p_4 ) = \langle \psi^* _{\mathbf p_1}\psi^* _{\mathbf p_2}\psi _{\mathbf p_3} \psi _{\mathbf p_4}\rangle,
\end{equation}
we can write the equation of motion as

\begin{equation}
 \frac{\D f}{\D t} (\mathbf x, \mathbf p)= 2 g \mathrm {Im} \int (\prod_{i=1} ^3\D \mathbf q_i) J_{\mathbf p \mathbf q_1 \mathbf q_2 \mathbf q_3} \delta(\mathbf p + \mathbf q_1-\mathbf q_2 -\mathbf q_3)
 \label{eq:WignerEvolution}
\end{equation}

Now under the assumption of an initial Gaussian distribution, the four-point function factors into a sum of two-point functions (the Wick contractions) at $t=0$. If we assume this holds at all times, rather than only at $t=0$, we obtain

\begin{equation}
    \frac{\D f}{\D t} (\mathbf x,\mathbf p)= n_\mathbf p \,\mathrm{Im} \int \D\mathbf q\, g n_\mathbf q,
    \label{eq:WignerEOM}
\end{equation}
where 
\begin{equation}
    n_\mathbf p = \langle \psi^* _\mathbf{p} \psi _\mathbf{p} \rangle
\end{equation}
is the momentum space density. which vanishes since the interaction $g$ is real. We see that under this assumption about the four-point function $J$, the derivative of the Wigner function vanishes, that is, the distribution appears constant.

In reality, the four-point correlator $J$ changes with time, and since the distribution does not remain Gaussian, the connected correlations become relevant \cite{Zakharov}. By applying the Gross-Pitaevsky equation (Eq. (\ref{eq:GPE})) to each of the four fields in in the four point correlator $J$, we obtain an evolution equation for $J$, in the same way that we obtained the evolution equation for the Wigner function in Eq. (5). We get
\begin{equation}
    \frac{\partial J_{\mathbf p_1 \mathbf p_2 \mathbf p_3 \mathbf p_4}}{\partial t} = i(\Delta \omega) J_{\mathbf p_1 \mathbf p_2 \mathbf p_3 \mathbf p_4} -ig A_{\mathbf p_ 1 \mathbf p_2 \mathbf p_3 \mathbf p_4},
    \label{eq:Jevolution}
\end{equation}
\begin{equation}
    \Delta\omega = \omega_{\mathbf p_1}+\omega_{\mathbf p_2}-\omega_{\mathbf p_3}-\omega_{\mathbf p_4},
\end{equation}
where $A_{\mathbf p_1 \mathbf p_2 \mathbf p_3 \mathbf p_4}$ is a six-point correlation function. This arises in the evolution equation of $J$ due to the nonlinear term in the Gross-Pitaevskii equation. Due to our initial conditions of a Gaussian distribution with random phases, this six-point function can be factored into its Wick contractions, with negligible connected component,
\begin{equation}
    A_{\mathbf p_1 \mathbf p_2 \mathbf p_3 \mathbf p_4}= n_{\mathbf p_3} n_{\mathbf p_4}(n_{\mathbf p_2} + n_{\mathbf p_1}) - n_{\mathbf p_1} n_{\mathbf p_2} (n_{\mathbf p_3}+n_{\mathbf p_4}).
\end{equation}
As with the four-point function $J$, the connected component of this six-point function becomes non-negligible as the system evolves. However, this process is suppressed by another factor of the interaction strength $g$ so we will neglect these corrections to lowest order in $g$, and simply assume that the six-point correlation $A$ is constant.

We can solve the evolution equation Eq. (\ref{eq:Jevolution}), obtaining
\begin{equation}
    J_{\mathbf p_1 \mathbf p_2 \mathbf p_3 \mathbf p_4} = B \exp(i\Delta\omega t) + g A_{\mathbf p_1 \mathbf p_2 \mathbf p_3 \mathbf p_4}/\Delta\omega.
    \label{eq:Jgrowth}
\end{equation}
We obtained this expression in order to estimate the rate of change of the Wigner function, and so we need to evaluate the integral in Eq. (\ref{eq:WignerEvolution}).

The first term in $J$ is an oscillating function. When $t$ is large, its contribution to the integral over momenta becomes negligible due to the cancellation of phases. Specifically, we need $t$ to be larger than the typical value of $(\Delta \omega)^{-1}$, and therefore larger than the typical value of $\omega_\mathbf p ^{-1}$. Since the axions exist in virialized miniclusters, the typical frequency of the field is defined by the virial velocity $v$ of the minicluster, namely
\begin{equation}
    \omega \sim mv^2.
\end{equation}
So provided the time is much greater than the typical timescale for the variation of the occupation numbers,
\begin{equation}
    t\gg \hbar/(mv^2),
    \label{eq:timescaleCondition}
\end{equation}
then the contribution from the oscillating term is negligible. 

Physically, the fourth-order correlations $J$ begin at zero since the initial distribution of axions in the minicluster is a distribution of Gaussian waves. Due to the nonlinear self-interactions, they cannot remain at zero but must grow as the self-interactions drive the system away from its initial Gaussian distribution. As we see in Eq. (\ref{eq:Jgrowth}), they grow and oscillate around their steady state value $g A/\Delta \omega$, on a timescale set by the typical energy of virialized axions in the minicluster, $mv^2$, where $v$ is the virial velocity. If we look at this process on a timescale in the regime of Eq.\ (\ref{eq:timescaleCondition}), then the dynamics of the fourth-order correlation function can be neglected, and we can assume that $J$ has a constant, nonzero value from the beginning of the condensation process. For a realistic minicluster, this period is much shorter than the relaxation timescale that follows from this assumption, so the approximation is consistent.

Substituting our solution for $J$ into Eq. (\ref{eq:WignerEvolution}) and neglecting the contribution from the oscillating term, we obtain
\begin{equation}\begin{aligned}
    \frac{\D f}{\D t} &= \pi g^2 \int (\prod_{i=1} ^3\D \mathbf q_i) \\ &\times\left[n_{\mathbf p_2} n_{\mathbf p_3}(n_{\mathbf p_1} + n_\mathbf p) - n_\mathbf p n_{\mathbf p_1} (n_{\mathbf p_2}+n_{\mathbf p_3}) \right]  \\
    &\times \delta(\mathbf p + \mathbf q_1 -\mathbf q_2 -\mathbf q_3)\delta(\omega_\mathbf p + \omega_{\mathbf p_1} -\omega_{\mathbf p_2} -\omega_{\mathbf p_3}).
\end{aligned}\end{equation}

From this kinetic equation, we can obtain a rate for the evolution of the Wigner function towards its steady-state. 
\begin{align}
    \frac{\D f}{\D t} &\sim \frac{\pi g^2 n^2}{\Delta \omega} f = f/\tau\\
    \tau &= \frac{mv^2}{\pi g^2 n^2}.
    \label{eq:tauself}
\end{align}
In particular, the relaxation rate scales with the square of the self-interaction strength. 
%obviously this is a problem since 10^10 s is not longer than the age of the universe... I think i will rewrite this to not cite the numerical value directly but instead just the ratio with tau_G
\section{Discussion}
In Ref. \cite{Relax}, we introduced a formalism for analyzing the relaxation of the scalar field dark matter into its condensed state when both self-interactions and gravity are present, by analysing the evolution of the Wigner function. We previously estimated the relaxation time due to self-interactions without accounting for the upper limit on the momentum occupation in the axion field imposed by the virial velocity. When this natural cut-off is taken into account, we see that the growth of the four-point correlations is suppressed, never growing past a steady state value. The resulting relaxation rate is proportional to $g^2$. Since the self-interactions are small, this is gives a much longer relaxation time than calculated in Ref. \cite{Relax}. The timescale estimated here agrees with the timescale for condensation due to self interaction found in the recent simulations in Ref. \cite{Xiaolong2021}.

When both gravitational and self-interactions are present the equation of motion for the Wigner function has the form
\begin{equation}
    \frac{\D f}{\D t} = \frac{f}{\tau_\lambda} + \frac{f}{\tau_G} = \frac{f}{\tau_\mathrm{tot}},
\end{equation}
where $\tau_\lambda$ is the relaxation timescale due to self-interactions alone, $\tau_G$ is the relaxation timescale due to gravity alone (calculated in the section above), and the combined timescale is 
\begin{equation}
    \tau_\mathrm{tot} = \frac{\tau_\lambda\tau_G}{\tau_\lambda + \tau_G}.
    \label{eq:CorrectTimescale}
\end{equation}
This simple form is seen in the Wigner function formalism. In particular, there is no term that drives the Wigner function of the form $\partial_t f \sim g G f$. Since the gravitational and self-interactions operate on different distance scales, such cross terms do not arise at lowest order in the interaction strengths $g$ and $G$, and the two interactions can be analyzed separately. 
In physical scenarios relevant for QCD or ultralight axions, $\tau_G\ll \tau_\lambda$, so the self-interactions do not significantly speed up the relaxation process.

\begin{figure}
\includegraphics[width=90mm]{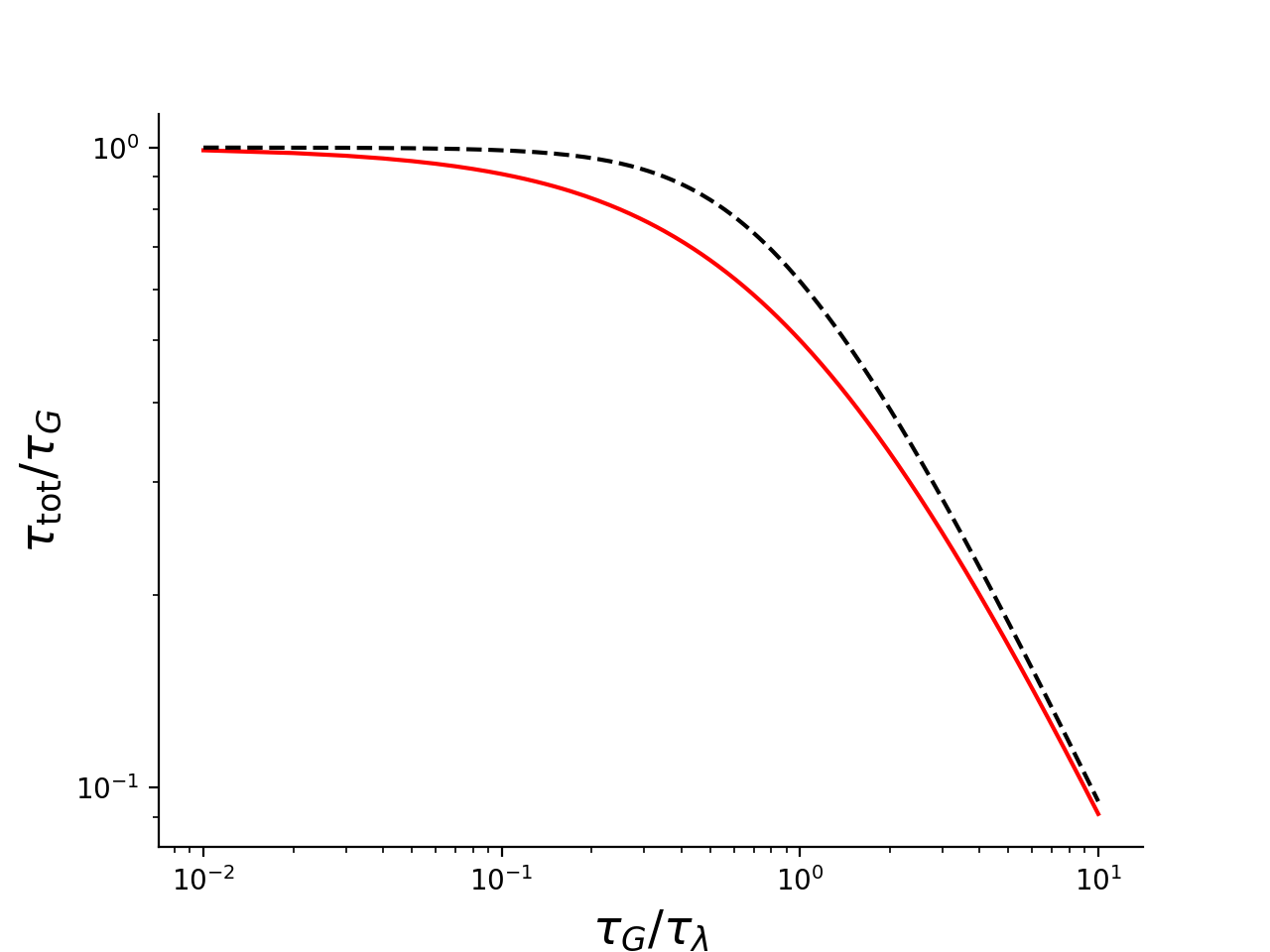}
\caption{Relaxation time when both self- and gravitational interactions are present, according to Eq.\ (\ref{eq:CorrectTimescale}) (solid red line) and Eq.\ (\ref{eq:IncorrectTimescale}) (dashed black line). For QCD axion, $\tau_G/\tau_L\ll 1$}
\label{fig:timescales}
\end{figure}

In Ref. \cite{Relax}, we previously predicted 
\begin{equation}
     \tau_\mathrm{tot} = \frac{2\tau_\lambda\tau_G}{\tau_\lambda+\sqrt{\tau_\lambda ^2 + 4\tau_G ^2}}.
     \label{eq:IncorrectTimescale}
\end{equation}
This equation occurred because the timescale for relaxation due to self-interaction arose out of a second-order equation for the Wigner function when we ignored the effect of the virial velocity in limiting the occupied momentum states in the axion field. When this effect is included, both self-interactions and gravity affect the evolution of the Wigner function at the same order, and the correct combined timescale is Eq.\ (\ref{eq:CorrectTimescale}). As shown in Fig.\ \ref{fig:timescales}, the two formulas approximate each other when either the relaxation time from self-interactions or from gravity greatly exceeds the other. While both expressions qualitatively fit the recent data obtained through numerical simulations in  Ref.\ \cite{Xiaolong2021}, we believe Eq.\ (\ref{eq:CorrectTimescale}) fits the data better than Eq. (\ref{eq:IncorrectTimescale}), since it is more appropriate to this physical situation and since it reduces the apparent systematic difference in the fit to the data in Ref.\ \cite{Xiaolong2021}.

Ref. \cite{Xiaolong2021} presented numerical simulations arguing that the correct time constant for condensation due to self-interactions scaled with $g^{-2}$. This work reinforces that result, and is consistent (up to an $\mathcal O(1)$ constant) with relaxation time caused by a self-interaction cross-section,
\begin{equation}
\tau_\lambda \sim b \frac{2 \sqrt{2} m^3 v^2}{3\pi^2 \sigma n^2},
\end{equation}
\begin{equation}
\sigma =\frac{ m^2 g^2}{2\pi},
\end{equation}
that was shown in the simulations and predicted in Ref. \cite{Levkov2018} within kinetic theory. Since these timescales for condensation into the Bose star are the scales associated with exponential growth of the correlations, they are only defined up to $\mathcal O(1)$.

Finally, we note that this analytic work confirms that for QCD axions and other scalar dark matter with weak self-interaction coupling strength, the relaxation timescale is set by gravity, since the corresponding gravitational timescale found in Ref. \cite{Levkov2018} and confirmed by simulations \cite{XiaolongSimulations} is:
\begin{equation}
    \tau_G=\frac{\sqrt{2}}{12\pi^3}\frac{mv^6}{G^2 n^2 \log(mv R)}.
\end{equation}
For QCD axions in miniclusters, this gives a ratio $\tau_\lambda/\tau_G \gtrsim 10^{12}$.

\section*{Acknowledgments}
\setlength{\parskip}{0pt}
We would like to thank Asimina Arvanitaki, Arka Banerjee, Noah Glennon, Sam McDermott, Nathan Musoke, Ethan Nadler, Tim M.P. Tait and Xiaolong Du for helpful conversations.  CPW would like to thank all workers who made this research possible, especially those at the University of New Hampshire (including Michelle Waltz and Katie Makem-Boucher). AEM's contributions to this project were supported by DOE Grant DE-SC0020220. This paper honors the memory of Breonna Taylor.

\bibliographystyle{apsrev4-1}
\bibliography{TonyAxion.bib}

\end{document}